\documentclass[12pt]{article}
\usepackage{color}
\usepackage{epsfig}
\usepackage{amsmath,amssymb}
\usepackage{amsthm}
\usepackage{setspace}
\usepackage{amsbsy}
\usepackage{natbib}
\usepackage{cases}
\usepackage{booktabs}
\usepackage{enumitem}
\usepackage{verbatim}
\usepackage{csquotes}

\newtheorem{theorem}{Theorem}

\newtheorem{lemma}{Lemma}
\newtheorem{definition}{Definition}
\newtheorem{property}{Property}

\DeclareMathOperator*{\argmin}{\arg\!\min}

\newcommand{\ra}[1]{\renewcommand{\arraystretch}{#1}}
\newcommand{\intvec}{{\boldsymbol \beta}_0^*}
\newcommand{\intscal}{\beta_0^*}
\newcommand{\vect}[1]{\boldsymbol{\mathbf{#1}}}

\newcommand{\intnostarvec}{{{\boldsymbol \beta}_0}}

\begin{document}
\renewcommand{\arraystretch}{1.5}
 \title{Quantile universal threshold for model selection}

\author{Caroline Giacobino\footnote{Department of Mathematics, University of Geneva; caroline.giacobino@unige.ch} \quad  \quad Sylvain Sardy\footnote{Department of Mathematics, University of Geneva; sylvain.sardy@unige.ch}\\
Jairo Diaz Rodriguez \footnote{Department of Mathematics, University of Geneva; jairo.diaz@unige.ch}  \quad  \quad 
Nick Hengartner \footnote{Theoretical Biology and Biophysics group, Los Alamos National Laboratory; nickh@lanl.gov}
}
\date{}

 \maketitle
\begin{quote}
Efficient recovery of a low-dimensional structure from high-dimensional data has been pursued in various settings including wavelet denoising, generalized linear models
and low-rank matrix estimation.
 By thresholding some parameters to zero, estimators such as lasso, elastic net and subset selection
 allow to perform not only parameter estimation but also variable selection, leading to sparsity.
Yet one crucial step challenges all these estimators: the choice of the threshold parameter~$\lambda$.
If too large, important features are missing; if too small, incorrect features are included.

Within a unified framework, we propose a new selection of $\lambda$ at the detection edge under the null model. To that aim, we introduce the concept of a zero-thresholding function and a null-thresholding statistic, that we explicitly derive for a large class of estimators.
The new approach has the great advantage of transforming the selection of $\lambda$ from an unknown scale to a probabilistic scale with the simple selection of a probability level.
Numerical results show the effectiveness of our approach in terms of model selection and prediction.

\bigskip
\footnotesize{Keywords: Convex optimization, high-dimensionality, sparsity, regularization, thresholding, variable screening.}

\end{quote}




\newpage

\section{Introduction}
\label{sct:intro}

Many real world examples in which 
the number of features $P$ of the model can be dramatically larger than the sample size~$N$ have been identified in various domains such as genomics, finance and image classification, to name a few.
In those instances, the maximum likelihood estimation principle  fails.  Beyond existence and 
uniqueness issues, it tends to perform poorly when $P$ is large relative to $N$ due to its high variance.  
Motivated by the seminal papers of \citet{Jame:Stei:esti:1961} and \citet{Tikhonov63},
a considerable amount of literature has concentrated 
on parameter estimation using regularization techniques.
In both parametric and nonparametric models, a reasonable prior or constraints are set on the parameters in order to reduce the variance of the estimator
and the complexity of the fitted model, at the price of a bias increase.

We consider a class of regularization techniques, called \emph{thresholding}, which:
\begin{enumerate}[label={(\roman*)}]
\item assumes a certain transform ${\boldsymbol \xi}^*=g({\boldsymbol \beta}^*)   \in {\mathbb R}^Q$ of the true model parameter  ${\boldsymbol \beta}^*   \in {\mathbb R}^P$  is \emph{sparse}, meaning
\begin{equation}\label{eq:Model*}
{\cal S}^*:=\{q\in\{1,\ldots,Q\}:\, \xi^*_q \neq 0\} 
\end{equation}
has small cardinality. For example, coordinate-sparsity is induced by $g({\boldsymbol \beta}^*)={\boldsymbol \beta}^*$,
whereas variation-sparsity is induced by $g({\boldsymbol \beta}^*)=B{\boldsymbol \beta}^*$ with $B$ the first order difference matrix;
 \item results in an estimated support
 \begin{equation}\label{eq:ModelHat}
 \hat {\cal S}_\lambda:=\{q\in\{1,\ldots,Q\}:\, \hat \xi_{\lambda,q} \neq 0\}
 \end{equation}
 whose cardinality is governed by the choice of a threshold parameter~$\lambda \geq 0$.
\end{enumerate}

Thresholding techniques  are employed in various settings such as
linear regression \citep{Dono94b,Tibs:regr:1996}, generalized linear models \citep{ParkHastie07},
 low-rank matrix estimation \citep{Mazumder:STMiss:2010,Cai:soft:2010},
density estimation \citep{CIS-130924,TVdensity10},
linear inverse problems \citep{DonohoWV95},
compressed sensing \citep{Donoho:CS:06,Candes:Romberg:2006}
and time series \citep{SardyNetoTseng12}.

Selection of the threshold is crucial to perform effective model selection. 
It amounts to selecting basis coefficients in wavelet denoising, or genes responsible for a cancer type in microarray data analysis.
In change-point detection, it is equivalent to detecting locations of jumps of a function. 
A too large $\lambda$ results in a simplistic model missing important features whereas a too small $\lambda$ leads to a model including many features outside the true model.
A typical goal is \emph{variable screening}, that is,
\begin{equation} \label{eq:VS}
\hat {\cal S}_\lambda \supseteq {\cal S}^* 
\end{equation}
holds with high probability,
along with
few false detections $\{q:\, \hat \xi_{\lambda,q} \neq 0, \  \xi^*_q=0 \}$. 
For a suitably chosen $\lambda$, certain estimators allow variable screening. 
The optimal threshold for  model identification often differs from the threshold aimed at prediction optimality
\citep{CIS-209692,CIS-220296,Mein:Bhlm:high:2006,Zou:adap:2006},
and it turns out that models aimed at good prediction are typically more complex. 

Classical methodologies to select $\lambda$ consist in minimizing a criterion. Examples include cross-validation, AIC~\citep{AkaikeAIC73}, BIC~\citep{Schw:esti:1978} and Stein unbiased risk estimation (SURE)~\citep{Stein:1981}.
In low-rank matrix estimation, \citet{Owen:cv:2009} and \citet{Josse11b} employ cross-validation whereas \citet{CandesSURE:2013} and \citet{JosseSardy2015} apply SURE.
The latter methodology is also used in regression  \citep{Dono94b,ZHT07,Tibs:Tayl:degr:2012}, and reduced rank regression
\citep{Mukherjee01062015}.
Because traditional information criteria do not adapt well to the high-dimensional setting, generalizations such as GIC \citep{FanTang13}
and EBIC \citep{ChenChen08} have been suggested.

\bigskip
In this paper, we propose a new threshold selection method that aims at a good identification of the support ${\cal S}^*$,
and that follows the same paradigm in various domains.
Our approach has the advantage of transforming the selection of $\lambda$ from an unknown scale to a probabilistic scale with the simple selection of a probability level.
In Section~\ref{subsct:regul}, we first review thresholding estimators in  linear regression, generalized linear models, low-rank matrix estimation and density estimation.
We then introduce the key concept of a zero-thresholding function in Section~\ref{sct:framework} and derive explicit formulations.
In Section~\ref{subsct:QUT}, we define the null-thresholding statistic, which leads to our proposal: the quantile universal threshold.
Some properties are derived.
Finally, we illustrate the effectiveness of our methodology in Section~\ref{sct:cm} with four real data sets and simulated data.
The appendices contain a proof, technical details and supplementary simulation studies.

\section{Review of thresholding estimators}
\label{subsct:regul}

Thresholding estimators are extensively used in the following domains.

\bigskip \noindent
{\bf Linear regression}. 
Consider the linear model
\begin{equation} \label{eq:regression}
  {\bf Y}=X_0 \intvec + X{\boldsymbol \beta}^{*}+\sigma {\boldsymbol \epsilon}, \quad {\boldsymbol \epsilon} \sim {\rm N}({\bf 0}, I_N),
\end{equation}
where $X_0$ and  $X$ are matrices of covariates or discretized basis functions of sizes $N \times P_0$ and $N \times P$ respectively, and
$\intvec$, ${\boldsymbol \beta}^*$ are unknown coefficients.
The vector $\intvec$ corresponds to $P_0$ parameters assumed a priori to be nonzero, as is the case for the intercept. 

For an observed ${\bf y}$, a large class of  estimators 
is of the form
\begin{equation} \label{eq:penalizedestimator}
(\hat \intnostarvec_\lambda, \hat {\boldsymbol \beta}_\lambda) \in \argmin_{(\intnostarvec, {\boldsymbol \beta})  \in \mathbb{R}^{P_0+P}} \, {\cal L}(X_0 \intnostarvec + X{\boldsymbol \beta},{\bf y}) + p_\lambda(g( {\boldsymbol \beta})),
\end{equation}
for a given loss ${\cal L}$  and function $g$.
A well chosen penalty $p_\lambda$ induces
sparsity in $\hat {\boldsymbol \xi}_\lambda= g(\hat {\boldsymbol \beta}_\lambda)$.
Note that the element notation ``$\in$''  indicates the minimizer might not be unique.
In the following, we assume for simplicity that $P_0=0$.
The lasso \citep{Tibs:regr:1996} 
\begin{equation} \label{eq:lasso}
\hat {\boldsymbol \beta}_\lambda^{\rm lasso} \in \,\argmin_{{\boldsymbol \beta} \in {\mathbb R}^P} \, \frac{1}{2} \|{\bf y}-X{\boldsymbol \beta}\|_2^2 + \lambda \|{\boldsymbol \beta}\|_1
\end{equation}
is among the most popular techniques.
Other examples include:
\begin{enumerate}[label={(\roman*)}]
 \item Total variation \citep{ROF92}, WaveShrink \citep{Dono94b}, adaptive lasso \citep{Zou:adap:2006}, group lasso \citep{Yuan:Lin:mode:2006},
  generalized lasso \citep{CIS-255162}, sparse group lasso \citep{SGlasso13},
 least absolute deviation (LAD) lasso  \citep{CIS-215377}, which minimizes $\|{\bf y}-X{\boldsymbol \beta} \|_1+\lambda \| {\boldsymbol \beta} \|_1$,
 square root lasso  \citep{BCW11}, which minimizes $\|{\bf y}-X{\boldsymbol \beta} \|_2+\lambda \| {\boldsymbol \beta} \|_1$
 and group square root lasso  \citep{BLS14}.
 \item Subbotin lasso \citep{SardySLIC09} where $p_\lambda({\boldsymbol \beta})=\lambda \| {\boldsymbol \beta} \|_\nu^\nu $, 
 $\nu \leq 1$, best subset selection, which is equivalent to Subbotin lasso with $\nu=0$,
 smoothly clipped absolute deviation (SCAD) \citep{Fan:Peng:PLS:2004}, minimax concave penalty (MCP)  \citep{CIS-251788} and smooth lasso \citep{SardySBITE2012}.

\end{enumerate}
Convex methodologies (i) also include the Dantzig selector \citep{Cand:Tao:dant:2007}.
Note that although ridge regression \citep{ridgeHK}, bridge \citep{Fu:1998} and smoothing splines \citep{Wahb:spli:1990} are of the form~\eqref{eq:penalizedestimator},
they do not threshold.




\bigskip \noindent
{\bf Generalized linear models} (GLMs).
The canonical model assumes the log-likelihood is of the form
\begin{equation} \label{eq:defh}
\ell \left( \intnostarvec,{\boldsymbol \beta}; {\bf y} \right)  =  \sum_{n=1}^{N} \left[  y_n \theta_n
- b  \left(\theta_n \right)   \right] \quad {\rm with} \quad
\theta_n={{\bf x}_0}_n^{\rm T} \intnostarvec + {\bf x}_n^{\rm T} {\boldsymbol \beta},
\end{equation}
$b$ a known function, ${{\bf x}_0}_n$ and ${\bf x}_n$ denoting the $n$th row of $X_0$ and $X$ respectively \citep{NW72}.
As an extension of lasso, \citet{SAT01} and \citet{ParkHastie07} define
\begin{equation} \label{eq:GLMlassowithA}
(  \hat \intnostarvec_\lambda, \hat {\boldsymbol \beta}_\lambda )  \in \,\argmin_{\left( \intnostarvec, {\boldsymbol \beta} \right)  \in \mathcal{F}} \,
- \ell \left( \intnostarvec,{\boldsymbol \beta}; {\bf y} \right) 
+ \lambda \| {\boldsymbol \beta} \|_1,
\end{equation}
where
$\mathcal{F}:=\left\lbrace  \left( \intnostarvec, {\boldsymbol \beta}\right)  \in \mathbb{R}^{P_0+P} \, | \, X_0 \intnostarvec + X {\boldsymbol \beta} \in {\Theta}^N \right\rbrace$
and ${\Theta}:=\{ \theta \in \mathbb{R} \, | \, b(\theta) < \infty \}$.
Other penalties such as group lasso \citep{Meier2008} have been proposed.

\bigskip \noindent
{\bf Low-rank matrix estimation}. Consider the model
$Y= X^* + \sigma Z$,
where  $X^*$ is a low-rank matrix and $Z_{ij} \stackrel{{\rm i.i.d.}} \sim {\rm N}(0,1)$.
Inspired by lasso, an estimate of  $X^*$  \citep{Mazumder:STMiss:2010,Cai:soft:2010} is given by
\begin{equation*} \label{eq:tracenorm}
 \argmin_{{X} \in \mathbb{R}^{N\times P}} \, \frac{1}{2} \|Y-X \|_F^2 + \lambda \| X \|_*,
\end{equation*}
where  $\| \cdot \|_F$  and $\| \cdot \|_*$ respectively denote the Frobenius and trace norm. 
For a fixed $\lambda$,
the solution is $\hat{X}=U {\rm diag}({\hat {{\bf d}}}_\lambda) V^{\rm T}$ with $Y=U {\rm diag}( {\bf d}) V^{\rm T}$ the singular value decomposition of $Y$,
and $\hat {d}_{{\lambda},i}=\max(d_i - \lambda, 0)$.


\bigskip \noindent
{\bf Density estimation}. Let $Y_1, \ldots, Y_N \stackrel{{\rm i.i.d.}}{\sim} \varphi$. 
A regularized estimate $\hat {\boldsymbol \varphi}_\lambda$ of the discretized density ${\boldsymbol \varphi}^*=[\varphi({\bf y}_{(1)}), \ldots, \varphi({\bf y}_{(N)}) ]$ is
$$
\hat {\boldsymbol \varphi}_\lambda = \argmin_{{\boldsymbol \varphi} \in {\mathbb R}^N}\, - \sum_{n=1}^N \log \varphi_n + \lambda  \|B {\boldsymbol \varphi} \|_1 \quad {\rm s.t.} \quad  {\boldsymbol a}^{\rm T} {\boldsymbol \varphi}=1,
$$
where $a_1=({\bf y}_{(2)}-{\bf y}_{(1)})/2$, $a_n=({\bf y}_{(n+1)}-{\bf y}_{(n-1)})/2$, $n=2,\ldots,N-1$, $a_N=({\bf y}_{(N)}-{\bf y}_{(N-1)})/2$, ${\bf y}_{(k)}$ denotes the $k$th order statistic and
$B$ is the first order difference matrix \citep{TVdensity10}.

\bigskip

Motivated by the preceding examples in GLMs and low-rank matrix estimation, a definition of a thresholding estimator is the following.
\begin{definition}\label{def:thresholding}
 Assume ${\bf Y} \sim f_{({\boldsymbol \eta}^*, {\boldsymbol \beta}^*)}$, with ${\boldsymbol \xi}^*=g ({\boldsymbol \beta}^*)$ sparse for a certain function~$g$
and ${\boldsymbol \eta}^*$ a vector of nuisance parameters.
 Let $\hat {\boldsymbol \beta}_\lambda({\bf Y})$ be an estimator indexed by $\lambda\geq 0$. We call
 $\hat {\boldsymbol \xi}_\lambda({\bf Y})=g \, \circ \, \hat {\boldsymbol \beta}_\lambda({\bf Y})$
 a \emph{thresholding estimator} 
 if $${\mathbb P}(\hat {\boldsymbol \xi}_\lambda({\bf Y})={\bf 0})>0
 \quad \mbox{ for some  finite } \lambda.
 $$
 \end{definition}
We make use of this definition when introducing the zero-thresholding function in the next section and our methodology in Section~\ref{sct:Defpro}. 

\section{The zero-thresholding function}
\label{sct:framework}
%

%


A key property shared by a class of estimators is to set the estimated parameters to zero for a sufficiently large but finite threshold $\lambda$. 
This leads to the following definition.



\begin{definition}\label{def:zerothresholdingfct}
 A thresholding estimator $\hat {\boldsymbol \xi}_\lambda({\bf Y})$  admits a zero-thresholding function 
 $\lambda_0({\bf Y})$ if
 \begin{equation*} \label{eq:zeroTF}
 \hat {\boldsymbol \xi}_\lambda({\bf Y}) = {\bf 0} \quad \Leftrightarrow \quad  \lambda \geq \lambda_0({\bf Y}) \quad  \mbox{ almost everywhere}.
  \end{equation*}
\end{definition}

%
The zero-thresholding function is hence determined uniquely up to sets of measure zero. Note that the equivalence implies equiprobability between setting all coefficients to zero and selecting the threshold large enough. 
It turns out that such a function has a closed form expression in many instances. 
Below we derive a catalogue 
for the estimators reviewed in Section~\ref{subsct:regul}. 

%


\bigskip \noindent
{\bf Linear regression}. Explicit formulations are the following:
\begin{itemize}
 \item Lasso, WaveShrink and the Dantzig selector: $ \lambda_0({\bf y})= \|X^{\rm T} {\bf y} \|_\infty$;  
 SCAD and MCP share the same zero-thresholding function when $X$ is orthonormal.
 For adaptive lasso, $\lambda_0({\bf y})= \|W X^{\rm T} {\bf y} \|_\infty$, where $W$ is a diagonal matrix of weights, 
 for LAD-lasso, $\lambda_0({\bf y})= \|X^{\rm T} {\rm sgn}({\bf y}) \|_\infty$, where ${\rm sgn}(\cdot)$ is the sign function applied componentwise,
 and for square root lasso, $\lambda_0({\bf y})=\| X^{\rm T} {\bf y} \|_\infty/\|{\bf y} \|_2$.
  \item Group lasso and square root lasso: if the parameters are partitioned into $G$ prescribed groups so that
 $p_\lambda({\boldsymbol \beta})=\lambda \sum_{g=1,\ldots,G} \| {\boldsymbol \beta}_g \|_2$, the zero-thresholding function
 is respectively $\lambda_0({\bf y})=\max_{g=1,\ldots,G}\|X_g^{\rm T} {\bf y} \|_2 $ and $\lambda_0({\bf y})=\max_{g=1,\ldots,G}\|X_g^{\rm T} {\bf y} \|_2/\|{\bf y} \|_2 $.
 \item Generalized lasso: Assuming $B$ has full row rank, let ${\cal I}$ denote a set of column indices such that $B_{{\cal I}}$, the submatrix of $B$ with columns indexed by ${\cal I}$, is invertible.
 Then, $\lambda_0({\bf y})=\| A_1^{\rm T} (I - P_{A_{2}}) {\bf y} \|_\infty$,
 where $P_X$ is the orthogonal projection matrix onto the range of~$X$,
 $A_1=X_{{\cal I}} B_{{\cal I}}^{-1}$, $A_2=X_{\bar {\cal I}}-X_{{\cal I}}B_{{\cal I}}^{-1} B_{\bar {\cal I}}$ and $\bar {\cal I}$ is the complement of ${\cal I}$.
In one-dimensional total variation, $\lambda_0({\bf y})= \| (BB^{\rm T})^{-1}B{\bf y}\|_\infty$.
 \item Best subset: 
  \begin{equation} \label{eq:BSS0thresh}
  \lambda_0({\bf y})= \max_{p=1,\ldots, {\rm rank}(X)} \frac{\Delta_p({\bf y})}{p},
  \end{equation} 
 where $\Delta_p({\bf y})= \frac{1}{2}\max_{\{ {\cal I}\subset \{1,\ldots,P \}:  |{\cal I}|=p \}} \| P_{X_{{\cal I}}} {\bf y} \|_2^2$.
 For $X$ orthogonal, $\lambda_0({\bf y})=\Delta_1({\bf y})$.
 \item Subbotin lasso: we conjecture that
 $$
  \lambda_0({\bf y})=\frac{2(1-\nu)}{(2-\nu)^2}  \max_{\{ {\cal I}\subset \{1,\ldots,P \}:\  1 \leq |{\cal I}| \leq {\rm rank}(X) \}} \frac{\| P_{X_{{\cal I}}} {\bf y} \|_2^2}{\| \hat {\boldsymbol \beta}_{\cal I}^{(p,\nu)} \|_\nu^\nu},
 $$
 where $\hat {\boldsymbol \beta}_{\cal I}^{(p,\nu)}=\frac{2(1-\nu)}{2-\nu} (X_{\cal I}^{\rm T}X_{\cal I})^{-1} X_{\cal I}^{\rm T}{\bf y}$ is the Subbotin-lasso estimate based on $X_{\cal I}$ for any $\nu \in [0,1)$.
 This expression simplifies to (\ref{eq:BSS0thresh}) if $\nu=0$, and to 
 $ \lambda_0({\bf y})= \{\|X^{\rm T}{\bf y}\|_\infty/(2-\nu)\}^{2-\nu} / \{ 2 (1-\nu) \}^{\nu-1}$ if $X$ is orthonormal.
\end{itemize}

For a convex objective function, derivation of the zero-thresholding function can be inferred from the Karush-Kuhn-Tucker conditions \citep{Rock}. As an example, we consider LAD-lasso. A given ${\boldsymbol \beta} \in \mathbb{R}^P$ is a minimum of the objective function $f({\boldsymbol \beta})=\|{\bf y}-X{\boldsymbol \beta} \|_1+\lambda \| {\boldsymbol \beta} \|_1$ if and only if ${\bf 0} \in \partial f({\boldsymbol \beta})$, the subdifferential of $f$ evaluated at ${\boldsymbol \beta}$.
The zero-thresholding function $\lambda_0({\bf y})= \|X^{\rm T} {\rm sgn}({\bf y}) \|_\infty$ then follows from the result for all ${\bf y} \in ({\mathbb{R}^*})^N$, $\partial f({\bf 0})=-X^{\rm T} {\rm sgn}({\bf y}) + \lambda \,{[-1,1]}^P$.

Such a derivation can also be performed for estimators with a composite penalty involving a two-dimensional parameter
${\boldsymbol \lambda}=(\lambda^{(1)}, \lambda^{(2)}$), for example:
\begin{itemize}
 \item Elastic net \citep{ZouHastie05} where $p_{{\boldsymbol \lambda}}({\boldsymbol \beta})=\lambda^{(1)} \|{\boldsymbol \beta} \|_1 + \lambda^{(2)} \|{\boldsymbol \beta}\|_2^2$:
 regardless of~$\lambda^{(2)}$, $\lambda^{(1)}_0({\bf y}; \lambda^{(2)})=\|X^{\rm T} {\bf y} \|_\infty$.
 \item Fused lasso \citep{CIS-191673}  where $p_{{\boldsymbol \lambda}}({\boldsymbol \beta})=\lambda^{(1)} \|{\boldsymbol \beta} \|_1 + \lambda^{(2)} \sum_{p=2}^P|\beta_{p}-\beta_{p-1}|$:
 assuming $X$ is orthonormal, $\lambda^{(1)}_0({\bf y}; \lambda^{(2)})=\|\hat {\boldsymbol \beta}_{(0,\lambda^{(2)})}({\bf y}) \|_\infty$.
\end{itemize}



\bigskip \noindent
{\bf Generalized linear models}. The following Lemma shows that 
although the lasso GLM solution defined in~\eqref{eq:GLMlassowithA} might not be unique, its fit is unique (see Appendix~\ref{app:prooflemma} for the proof).

\begin{lemma}\label{uniquefit}
Assume $b$ is strictly convex on $\Theta$. For any fixed $X_0$, $X$, $\vect{y}$ and $0 \leq \lambda < \infty$, $X_0 \hat {{\boldsymbol \beta}_0}_{\lambda} + X \hat{\boldsymbol{\beta}}_\lambda$ is unique. 
\end{lemma}

The zero-thresholding function of $\hat {\boldsymbol \beta}_\lambda$ is given in~\eqref{cor:zerothresholdingfctlassoGLM} below.
Its derivation is based on  Theorem~\ref{thm:KKT} whose proof can be found in Appendix~\ref{app:proof}.

\begin{theorem}\label{thm:KKT}
Assume $b$ in~\eqref{eq:defh} is convex on ${\Theta}$ open, and let \\
${\boldsymbol \mu}(\intnostarvec) = (b^\prime({{\bf x}_0}_1^{\rm T} \intnostarvec),\ldots,
b^\prime({{\bf x}_0}_N^{\rm T} \intnostarvec))^{\rm T}$. For any fixed $X_0$, $X$, ${\bf y}$ and $0 \leq \lambda < \infty$,

\begin{equation*}
{( \hat \intnostarvec_\lambda,{\boldsymbol 0} )  \in \,\argmin_{\left( \intnostarvec,{\boldsymbol \beta}\right)  \in \mathcal{F}} \, 
-\ell \left( \intnostarvec,{\boldsymbol \beta}; {\bf y}\right) 
+ \lambda \| {\boldsymbol \beta} \|_1
\iff}
\begin{cases}
X_0 \hat \intnostarvec_\lambda \in {\Theta}^N \\
{X_0}^{\rm T} {\bf y} = {X_0}^{\rm T} {\boldsymbol \mu}  ( \hat \intnostarvec_\lambda ) \\
\| X^{\rm T} [ {\bf y} - {\boldsymbol \mu} ( \hat \intnostarvec_\lambda) ]  \|_{\infty} \leq \lambda.\,
\end{cases}
\end{equation*}
\end{theorem}

\bigskip

Hence, for a strictly convex $b$ and setting $\hat {\boldsymbol \beta}_\lambda={\bf 0}$ if $\lambda=+\infty$,
the zero-thresholding function is
\begin{equation} \label{cor:zerothresholdingfctlassoGLM}
\lambda_0({\vect{ y} })= 
\begin{cases}
\| X^{\rm T} [ {\vect{ y} } - {\boldsymbol \mu}  ( \vect{v} ) ]  \|_{\infty} & \mbox{if} \ {\vect{ y} } \in {\cal D},\\
+ \infty & \mbox{otherwise,}
\end{cases}
\end{equation}
with $\vect{v}$ any vector such that
\begin{equation}
\begin{cases}\label{GLMKKT}
X_0 \vect{v} \in {\Theta}^N \\
{X_0}^{\rm T} {\vect{ y} } = {X_0}^{\rm T} {\vect{ \mu} }  ( \vect{v}) 
\end{cases}
\end{equation}
and ${\cal D}= \{\vect{y} \mid \, \exists \vect{v} \in \mathbb{R}^{P_0}\mbox{ solution to~\eqref{GLMKKT}} \}$. 

For the group lasso GLM, one obtains similarly
\begin{equation*}\label{grlassoGLMzero}
\lambda_0({\vect{ y} })= 
\begin{cases}
\max_{g=1,\ldots,G}  \| X_{g}^{\rm T} [ {\vect{ y} } - {\boldsymbol \mu}  ( \vect{v} ) ]  \|_{2} & \mbox{if} \ {\vect{ y} } \in {\cal D},\\
+ \infty & \mbox{otherwise.}
\end{cases}
\end{equation*}

Lemma~\ref{uniquefit} implies $\lambda_0({\vect{ y} })$ does not depend on which solution $\vect{v}$ to~\eqref{GLMKKT} is chosen.
The set ${\cal D}$ is the set of values based on which the maximum likelihood estimate (MLE) of $({{\boldsymbol \beta}_0},{\boldsymbol \beta})$ with constraint $\hat {\boldsymbol \beta}=0$ exists.
If the response variable is Gaussian, note that ${\cal D}={\mathbb R}^N$.
An explicit formulation of ${\cal D}$ when the intercept is unpenalized ($X_0={\bf 1}$) is given in Table~\ref{tab:somedistri}.
For an arbitrary matrix $X_0$ and under certain assumptions, \citet{Giacobinothesis} shows that ${\cal D}$ coincides with the set of values $\vect{ y}$ such that lasso GLM admits a solution.
In particular, the following property holds.
\begin{property} Consider a Poisson, logistic or multinomial logistic regression model. Then, for any fixed $X_0$, $X$ and $0 < \lambda < \infty$, and any observed value $\vect{ y}$, lasso GLM defined in~\eqref{eq:GLMlassowithA} admits a solution if and only if a MLE of $({{\boldsymbol \beta}_0},{\boldsymbol \beta})$ with constraint $\hat {\boldsymbol \beta}=0$ exists.
\end{property}

\begin{table}[h!]
\footnotesize
\caption{Values of $b^\prime({\intscal})$, ${\mathcal D}$ and ${\mathbb P}({\bf Y} \in {\cal D})$  when $X_0={\bf 1}$.}
\label{tab:somedistri}
\centering
\begin{tabular}{l|ccc} \hline
 Response distribution & $\mu=b^\prime({\intscal})$ & ${\mathcal D}$      & ${\mathbb P}({\bf Y} \in  {\cal D})$ \\ \hline \hline
 Gaussian     & $\intscal $  &  $\mathbb{R}^N$    & 1 \\
 Poisson      & $\exp\left( \intscal \right)$          & $\mathbb{N}^N \setminus \left\lbrace  {\bf 0} \right\rbrace $ & $1 - \exp\left( -N \mu  \right) $ \\ 
 Bernoulli    & $\exp\left( \intscal  \right) /\left( 1+\exp\left( \intscal\right) \right)$   & $\left\lbrace 0,1 \right\rbrace ^N \setminus \left\lbrace  {\bf 0}, {\bf 1}\right\rbrace $ & $1 - \mu^N-\left( 1-\mu \right) ^N$ \\ 
$ \operatorname {Binomial} \left( m, p \right)/m$     & $ \exp\left( \intscal  \right) /\left( 1+\exp\left( \intscal\right) \right)$ & $\left\lbrace 0,1/m,\ldots,1 \right\rbrace ^N \setminus \left\lbrace  {\bf 0}, {\bf 1}\right\rbrace $ & $1 - (\mu)^{mN}-\left( 1-\mu \right) ^{mN}$ \\ \hline
\end{tabular}
\end{table}
\normalsize

\bigskip \noindent
{\bf Low-rank matrix estimation}. The zero-thresholding function is $\lambda_0(Y)= \| {\bf d}\|_\infty$, the largest singular value of the noisy matrix~$Y$.

\bigskip \noindent
{\bf Density estimation}. The zero-thresholding function is $\lambda_0({\bf y})=\| {\bf w}\|_\infty$ with $w_k=N \sum_{i=1}^k a_{i} - k \sum_{i=1}^N a_i$, $k=1,\ldots,N-1$.

\section{The quantile universal threshold}
\label{subsct:QUT}

\subsection{Thresholding under the null}
\label{sct:Defpro}

Inspired by \citet{Dono94b},
we now consider the idea of choosing a threshold based on the null model ${\boldsymbol \xi}^* = {\bf 0}$, 
that is, selecting a threshold $\lambda$ such that $\{\hat {\boldsymbol \xi}_\lambda={\bf 0}\mid {\boldsymbol \xi}^*={\bf 0}\}$ holds with high probability.
From Definition~\ref{def:zerothresholdingfct}, the events $\{\hat {\boldsymbol \xi}_\lambda={\bf 0}\}$ and $\{\lambda \geq \lambda_0({\bf Y})\}$ are equiprobable.
This conducts us to the zero-thresholding function under the null model.


\begin{definition}\label{def:nullthresholdingstat}
 Assume $\hat {\boldsymbol \xi}_\lambda({\bf Y})$  admits a zero-thresholding function $\lambda_0({\bf Y})$.
 The  null-thresholding statistic  is
 \begin{equation} \label{eq:nullTS}
 \Lambda:=\lambda_0({\bf Y}_0)
  \end{equation}
   with ${\bf Y}_0=_d {\bf Y}$ under  $H_0: {\boldsymbol \xi}^*={\bf 0}$.
\end{definition}

Given a thresholding estimator and its null-thresholding statistic, selecting $\lambda$ large enough such that  ${\boldsymbol \xi}^*$ is recovered with probability $1-\alpha$
under the null model ${\boldsymbol \xi}^*={\bf 0}$ leads to the following new selection rule.

\begin{definition}\label{def:QUT}
The quantile universal threshold $\lambda^{{\rm QUT}}$ 
is the upper $\alpha$-quantile of $\Lambda$ defined in~\eqref{eq:nullTS}.
\end{definition}
 

We discuss the selection of $\alpha$ in Section~\ref{subsct:QUTprop}. As we will see, it turns out such a choice results in good empirical and theoretical properties even in the case ${\boldsymbol \xi}^* \neq {\bf 0}$.

If the distribution of $\Lambda$ is unknown, $\lambda^{\rm QUT}$ can be computed numerically by Monte Carlo simulation.
For instance, one can easily simulate realizations of square root lasso's null-thresholding statistic  $\Lambda=\| X^{\rm T} {\bf Y}_0\|_\infty/\| {\bf Y}_0\|_2$,  
and compute $\lambda^{\rm QUT}$ by taking the appropriate upper quantile.
Section~\ref{subsct:QUTprop} considers situations where a closed form expression of $\lambda^{\rm QUT}$ can be derived.

With the quantile universal threshold, selection of the regularization parameter is now redefined on a probabilistic scale through the probability level~$\alpha$. 
QUT is a selection rule designed for model selection as it aims at good identification of the support of the estimand ${\boldsymbol \xi}^*$.
If one is instead interested in good prediction, then the sparse model identified by QUT can be refitted by maximum likelihood.
Such a two step approach has been considered (see \citet{BuhlGeer11,ABVC13}) to mimic the behavior of adaptive lasso \citep{Zou:adap:2006}
and results in a smaller amount of shrinkage and bias of large coefficients.

\subsection{Instances of QUT}
\label{subsct:thrnull}

A QUT-like selection rule supported by theoretical results has appeared in the following three settings.

%

\bigskip\noindent
{\bf Wavelet denoising}. \citet{Dono94b} and \citet{Dono95asym} consider an orthonormal $P\times P$ wavelet matrix and select 
the threshold of soft-WaveShrink as $\lambda_P^{\rm universal}=\sigma \sqrt{2 \log P}$.
Under the null model with wavelet coefficients ${\boldsymbol \beta}^*={\bf 0}$, 
$\mathbb{P}(\hat {\boldsymbol \beta}_{\lambda_P^{\rm universal}}={\bf 0})
 \stackrel{P \rightarrow \infty}{\longrightarrow} 1$.  
It turns out that an oracle inequality and minimax properties  hold with $\lambda=\lambda_P^{\rm universal}$ over a wide class of functions, that is, when ${\boldsymbol \beta}^* \neq {\bf 0}$.
We show below that $\lambda_P^{\rm universal}=\lambda^{\rm QUT}$ for a small $\alpha$ tending to zero with $P$.

\bigskip
\noindent {\bf Linear regression}. Desirable properties of estimators such as the lasso, group lasso, square root lasso, group square root lasso or the Dantzig selector
are satisfied if the tuning parameter is set to $\lambda=c \lambda^{(0)}$ for a certain $c \geq 1$, such that the event
$\{\hat {\boldsymbol \beta}_{\lambda^{(0)}}={\bf 0} \mid {\boldsymbol \beta}^*={\bf 0} \}$ holds with high probability, for instance with $\lambda^{(0)}=\lambda^{\rm QUT}$ for a small $\alpha$.
More precisely, upper bounds on the estimation and prediction error, as well as the screening property~\eqref{eq:VS} hold with high probability
assuming certain conditions on the regression matrix, the support ${\cal S}^*$ of the coefficients and their magnitude; see \citet{BuhlGeer11,BCW11,BLS14} and references therein.

\bigskip\noindent
{\bf Low-rank matrix estimation}.   
Under the null model $X^*=0_{N \times P}$, it can be shown that with a noise level of $1/\sqrt{N}$, the empirical distribution of the singular values of the response matrix 
converges to a compactly supported distribution.  By setting any singular value smaller than the upper bound of the support to zero, \citet{OptiShrinkSV14} derive optimal singular value thresholding operators.




\bigskip
In these three settings, the importance of the null model to select the threshold or to derive theoretical properties is worth noticing.

\subsection{Properties of QUT}
\label{subsct:QUTprop}

Before considering the choice of $\alpha$ and deriving an explicit formulation of the quantile universal threshold in some settings, more theoretical properties are derived.
Upper bounds on the estimation and prediction error of the lasso tuned with $\lambda^{{\rm QUT}}$ as well as a sufficient condition for the screening property~\eqref{eq:VS} follow from the next property. 
\begin{property}
\bigskip Assume the $(L,{\cal S}^*)$-compatibility condition is satisfied for ${\cal S}^*$ of cardinality $s^*$ with $L=(\lambda + \lambda^{(0)}) / (\lambda-\lambda^{(0)})$, for a certain $\lambda^{(0)}$, $0<\lambda^{(0)}<\lambda$, that is,
$$
\phi_{\rm{comp}}(L,{\cal S}^*):=\min \left\lbrace  \,  \sqrt{s^*} \| X{\boldsymbol \beta} \|_2 /   \| {\boldsymbol \beta}_{{\cal S}^*}\|_1 \, \mid \, \| {\boldsymbol \beta}_{\bar {{\cal S}^*}}\|_1 \leq L \| {\boldsymbol \beta}_{{\cal S}^*}\|_1, \, {\boldsymbol \beta} \neq {\bf 0} \, \right\rbrace  >0.
$$ 
Then lasso~\eqref{eq:lasso} with $\lambda=\lambda^{{\rm QUT}}$ satisfies with probability at least $1-\alpha-{\mathbb P}( \lambda^{(0)} \leq \Lambda \leq \lambda)$ 
\begin{enumerate}[label={(\roman*)}]
\item \label{1} $\|X({\hat{\boldsymbol \beta}}_{\lambda} - {\boldsymbol \beta^*})   \|_2^2/2 \leq 8 (\lambda + \lambda^{(0)})^2  s^* / \phi^2_{\rm{comp}}(L,{\cal S}^*)$,
\item \label{2} $\| ({\hat{\boldsymbol \beta}}_{\lambda} - {\boldsymbol \beta^*})_{{\cal S}^*} \|_1 \leq A, \quad A=4 (\lambda + \lambda^{(0)}) s^* / \phi^2_{\rm{comp}}(L,{\cal S}^*)$,
\item \label{3} $\| ({\hat{\boldsymbol \beta}}_{\lambda})_{\bar{{\cal S}^*}} \|_1 \leq 4 s^* (\lambda + \lambda^{(0)})^2/\{ (\lambda -\lambda^{(0)}) \phi^2_{\rm{comp}}(L,{\cal S}^*)\} $.
\end{enumerate}
If, in addition,
\begin{equation*}
\min_{p \in {\cal S}^*} \, |{\boldsymbol \beta}_p^*| > A,
\end{equation*}
then with the same probability
\begin{equation*}
\hat {\cal S}_\lambda \supseteq {\cal S}^*.
\end{equation*}
\end{property}
Remark that ${\mathbb P}( \lambda^{(0)} \leq \Lambda \leq \lambda)$ can be made arbitrarily small for a well-chosen $\lambda^{(0)}$ as long as the $(L,{\cal S}^*)$-compatibility condition is met. The proof of the property is omitted as it is essentially the same as for Theorem~6.1 in \citet{BuhlGeer11}
using the fact that the key statistic they bound with high probability is the null-thresholding statistic
$\Lambda=\| X^{\rm T} {\boldsymbol \epsilon} \|_\infty=\lambda_0({\bf Y}_0)$ defined in~\eqref{eq:nullTS}. Note that the screening property is a direct consequence of \ref{2}.
Similar results can be shown for the group lasso, square root lasso, group square root lasso and the Dantzig selector.



Another important property of our methodology concerns the familywise error rate. Recall that when performing multiple hypothesis tests, it is defined as the probability of incorrectly rejecting at least one null hypothesis. In the context of variable selection, it is the probability of erroneously selecting at least one variable. It can be shown that if the null model is true, the familywise error rate is equal to the false discovery rate defined in Section~\ref{subsct:GaussianSimu}. Hence, Definition \ref{def:QUT} implies the following property.
\begin{property}
Any thresholding estimator tuned with $\lambda^{{\rm QUT}}$ controls the familywise error rate as well as the false discovery rate at level $\alpha$ in the weak sense.
\end{property}


The probability of the previous properties is determined by $\alpha$; we recommend $\alpha=0.05$ as 
\citet{BCW11}.
An alternative is to set $\alpha_P$ tending to zero as the number $P$ of covariates goes to infinity. \citet{Dono94b} implicitly select a rate of convergence of $\alpha_P=O(1/\sqrt{\log P})$ (\citet{JosseSardy2015} also select this rate). 

Finally, an explicit formulation of the quantile universal threshold can be derived in the following settings:

\begin{enumerate}[label={(\roman*)}]
\item In orthonormal regression with best subset selection and threshold $\sigma \sqrt{2 \log P}$ discussed in Section~\ref{subsct:thrnull}, the equivalent penalty is
$\lambda^{\rm QUT}=2 \lambda^{\rm BIC}=\sigma^2 \log P$ satisfying
$\bar F_\Lambda(\lambda^{\rm QUT}) \sim 1/\sqrt{\pi \log P}$.
This result can be inferred from  
the null-thresholding statistic $\Lambda=_d \|{\bf Z} \|_\infty^2/2$ using~\eqref{eq:BSS0thresh}, where $Z_i \stackrel{{\rm i.i.d.}} \sim {\rm N}(0,\sigma^2)$.
Generalizations such as GIC and EBIC also select a larger tuning parameter than BIC which performs poorly in the high-dimensional setting.

 
\item 
In total variation, the null-thresholding statistic converges in distribution to the
infinite norm of a Brownian bridge, leading to $\lambda^{\rm QUT}= \sigma \sqrt{P \log \log P}/2$ for $\alpha_P=O(1/\sqrt{\log P})$  \citep{SardyTseng04}.
For block total variation, the null-thresholding statistic tends to the maximum of a Bessel bridge, which distribution is
known \citep{PitmanYor99}.
 \item In group lasso with orthonormal groups, each of size $Q$, extreme value theory leads to
 $\lambda^{\rm QUT}=\sigma \sqrt{2 \log P + (Q-1) \log \log P - 2 \log \Gamma(Q/2)}$ 
\citep{SardySBITE2012}.
\end{enumerate}

\section{Numerical results of lasso GLM}
\label{sct:cm}

The QUT methodology for lasso and square root lasso is implemented 
in the {\bf qut} package which is available from the Comprehensive \textsf{R} Archive Network (CRAN).
%
%
In the following, QUT$_{\rm lasso}$ and QUT$_{\sqrt{{\rm lasso}}}$ stand for QUT applied to lasso and square root lasso respectively. CVmin refers to cross-validation,
CV1se to a conservative variant of CVmin which takes into account the variability of the cross-validation error \citep{cart84},
SS to stability selection \citep{stabsel10}
and GIC to the generalized information criterion \citep{FanTang13}.
When applying GIC and QUT$_{\rm lasso}$, the variance is estimated with \eqref{eq:sigma12} and \eqref{eq:sigma2QUT} respectively.
The level $\alpha$ is set to $0.05$.

\subsection{Real data}
\label{sct:data}

\begin{figure}[!ht]
   \begin{center}
    \includegraphics[width=\textwidth]{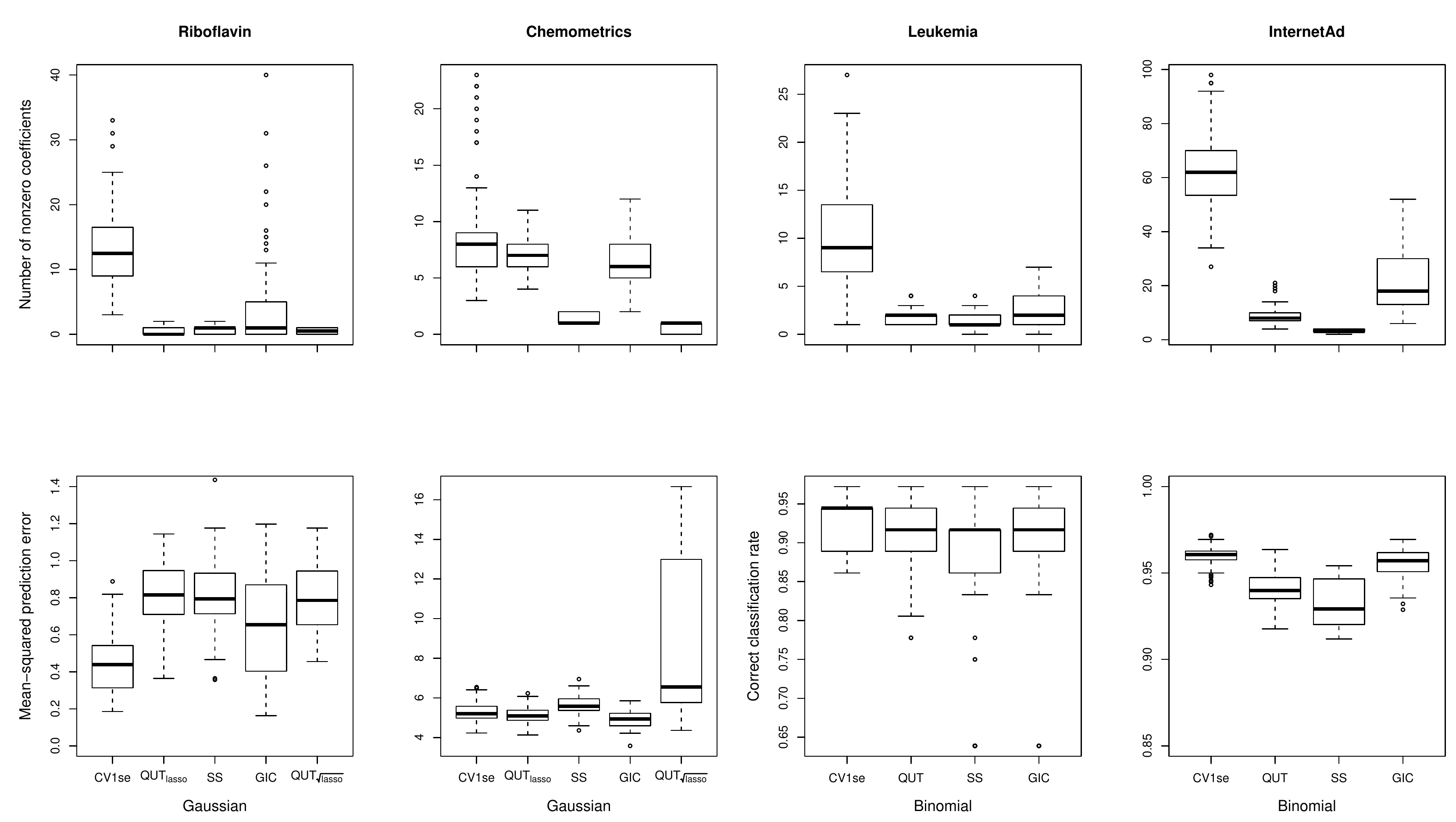}
   \end{center}
   \caption{Monte Carlo simulation based on four data sets: {\tt riboflavin} (Gaussian), {\tt chemometrics} (Gaussian), {\tt leukemia} (Binomial) and {\tt internetAd} (Binomial).
   We report the boxplots of the following statistics: the number of nonzero coefficients obtained from the training sets (top);
   the test set mean-squared prediction error for Gaussian responses and the correct classification rate for binomial responses (bottom).
      \label{fig:realdata}}
 \end{figure}

We  briefly describe the four data sets considered to illustrate our approach in Gaussian and logistic regression:


\begin{itemize}
 \item {\tt riboflavin}  \citep{PeterBulbiology:14}: Riboflavin production rate measurements from a population of Bacillus subtilis with sample size $N=71$ and expressions from $P=4088$ genes. 
 \item {\tt chemometrics} \citep{SardyISI08}: Fuel octane level measurements with sample size $N=434$ and $P=351$ spectrometer measurements. 
 \item {\tt leukemia} \citep{Golubetal1999}: Cancer classification of human acute leukemia cancer types based on $N=72$ samples of $P=3571$ gene expression microarrays. 
 \item {\tt internetAd} \citep{Kushmerick:1999:LRI:301136.301186}: Classification of $N=2359$ possible advertisements on internet pages based on $P=1430$ features. 
\end{itemize}
We randomly split one hundred times each data set into a training and a test set of equal size. 
Five lasso selection rules are compared including QUT. 
Except for CV1se, the final model is fitted by MLE with the previously selected covariates in order to improve prediction.
In Figure~\ref{fig:realdata},  we report the number of nonzero coefficients selected on the training set, as well as the test set mean-squared prediction error and correct classification rate.

Good predictive performance is achieved by QUT$_{\rm lasso}$ as well as GIC with a median model complexity between SS and CV1se.
QUT$_{\rm lasso}$ works remarkably well for {\tt chemometrics} and {\tt leukemia}.
By selecting a large number of variables CV1se results in efficient prediction, whereas SS and $\sqrt{{\rm lasso}}$ show poor predictive performance due to the low complexity of the model.
Moreover, GIC  exhibits a larger variability than QUT$_{\rm lasso}$ and QUT$_{\sqrt{{\rm lasso}}}$  in terms of number of nonzero coefficients.

%

\subsection{Synthetic data}
\label{subsct:GaussianSimu}

Two prominent quality measures of model selection are the true positive rate
${\rm TPR}:={\mathbb E}[ {\rm TPr}]$ and the false discovery rate ${\rm FDR}:={\mathbb E}[ {\rm FDr}]$, 
where ${\rm TPr}:=| \hat {\cal S}_\lambda \cap {\cal S}^*|/|{\cal S}^*|$, the proportion of selected nonzero features among all nonzero features,
 and ${\rm FDr}:=|\hat {\cal S}_\lambda \cap \bar {\cal S}^*|/|\hat {\cal S}_\lambda|$, the proportion of falsely selected features among all selected features.

\begin{table}[!ht]
\centering
\caption{Estimated TPR/FDR/RMSE based on the simulation of Section~\ref{subsct:GaussianSimu}.} 
\label{tab:tprfdr}
\centering
\bigskip
\ra{1.0}
\footnotesize
\begin{tabular}{rcccccc}
  \hline
  Method && \multicolumn{5}{c}{Response variable distribution}\\
 \cline{3-7}
 &&\multicolumn{1}{c}{Gaussian} && \multicolumn{1}{c}{Binomial} && \multicolumn{1}{c}{Poisson}          \\
 \hline
 &\emph{($\theta$,$\omega$,snr)} & \emph{(0.5,0,1)}   && \emph{(0.5,0,10)}   && \emph{(0.5,0,0.5)}   \\
 \hline
 {\bf lasso} \\
  CV1se &  									& 0.23/0.27/0.87 	&  	& 0.26/0.42/0.10	&  	& 0.28/0.34/3.35 \\ 
  QUT$_{\rm lasso}$	 &  					& 0.09/0.02/0.85 	& 	& 0.10/0.02/0.10	&  	& 0.37/0.57/2.94 \\ 
  SS &  									& 0.12/0.03/0.81 	&  	& 0.11/0.03/0.10  &  & 0.13/0.02/3.27 \\ 
  GIC &  									& 0.10/0.04/0.85 	&  	& 0.13/0.12/0.10  &  & 0.35/0.50/2.98  \\
  $\sqrt{\text{{\bf lasso}}}$\\ 
  QUT$_{\sqrt{{\rm lasso}}}$ &  			& 0.05/0.01/0.92    &  	&        			&  	&         \\ 
   \\
 &\emph{($\theta$,$\omega$,snr)} 			& \emph{(0.1,0,1)}	&	& \emph{(0.1,0,10)}	&	& \emph{(0.3,0,0.5)}   \\
 \hline
 {\bf lasso} \\
  CV1se &  									& 0.70/0.25/0.57 	&  	& 0.83/0.50/0.06 	&  	& 0.62/0.38/2.51 \\ 
  QUT$_{\rm lasso}$ &  						& 0.61/0.00/0.35 	&  	& 0.67/0.00/0.04 	&  & 0.64/0.44/1.96 \\ 
  SS &  									& 0.66/0.03/0.31 	&  	& 0.74/0.01/0.04 	&  & 0.40/0.02/2.12 \\ 
  GIC &  									& 0.68/0.13/0.36 	&  	& 0.78/0.13/0.04 &  & 0.64/0.47/2.03 \\ 
  $\sqrt{\text{{\bf lasso}}}$\\
  QUT$_{\sqrt{{\rm lasso}}}$ 			&  	& 0.24/0.00/0.80 	&  	&     				&  	&        \\ 
   \\
 &\emph{($\theta$,$\omega$,snr)} 			& \emph{(0.5,0.4,1)}&	&\emph{(0.5,0.4,10)}&	& \emph{(0.5,0.4,0.5)} \\
 \hline
 {\bf lasso} \\
  CV1se &  									& 0.18/0.79/0.67 	&  	& 0.15/0.80/0.08 &  & 0.24/0.82/2.56 \\ 
  QUT$_{\rm lasso}$ 					&  	& 0.13/0.71/0.63 	&  	& 0.12/0.78/0.09 &  & 0.26/0.82/2.41  \\ 
  SS &  									& 0.03/0.03/0.92 	&  	& 0.02/0.08/0.11 &  & 0.03/0.03/3.64 \\ 
  GIC &  									& 0.06/0.37/0.83 	&  	& 0.06/0.48/0.10  &  & 0.24/0.81/2.37 \\ 
  $\sqrt{\text{{\bf lasso}}}$\\
  QUT$_{\sqrt{{\rm lasso}}}$ &  			& 0.02/0.25/0.92 	&  &     &  &        \\ 
   \\
 &\emph{($\theta$,$\omega$,snr)} 			& \emph{(0.5,0,10)}  &&  \emph{(0.5,0,20)}   && \emph{(0.5,0,2)}   \\
 \hline
 {\bf lasso} \\
  CV1se &  									& 0.76/0.61/0.39 	&  	& 0.33/0.50/0.07 &  & 0.58/0.73/10.78 \\ 
  QUT$_{\rm lasso}$ &  						& 0.20/0.00/0.66 	&  	& 0.12/0.02/0.07 &  & 0.64/0.77/\phantom{0}9.04  \\ 
  SS &  									& 0.26/0.00/0.59 	&  	& 0.14/0.02/0.07 &  & 0.11/0.14/12.14 \\ 
  GIC &  									& 0.55/0.22/0.42 	&  	& 0.18/0.15/0.07 &  & 0.65/0.78 /\phantom{0}9.10  \\ 
  $\sqrt{\text{{\bf lasso}}}$\\
  QUT$_{\sqrt{{\rm lasso}}}$ 			&  	& 0.06/0.00/0.88 &  &    &  &     \\ 
   \hline
\end{tabular}
\end{table}

We perform a simulation based on \citet{ReidTibshFriedarchiv14}. Responses are generated from the linear, logistic and Poisson regression model with a sample size of $N=100$ and $P=1000$ covariates.
The intercept is set to one and unit noise variance is assumed in linear regression.
The true parameter ${\boldsymbol \beta}^*$ and predictor matrix $X$ are obtained as follows:
\begin{itemize} 
\item Elements of $X$ are generated randomly as $X_{ij} \sim {\rm N}(0, 1)$ with correlation between columns set to $\omega$.
\item The support of ${\boldsymbol \beta}^*$ is of cardinality $s^*=\lceil N^{\theta} \rceil$ and selected uniformly at random. Entries are generated from a ${\rm Laplace}(1)$ distribution and scaled according to a certain signal to noise ratio,
${\rm snr}={{\boldsymbol \beta}^*}^{\rm T}\Sigma_\omega {\boldsymbol \beta}^*$, $\Sigma_\omega$ being the covariance matrix of a single row of $X$ and for a noise variance $\sigma^2=1$ in the Gaussian case. 
\end{itemize}



Table~\ref{tab:tprfdr} contains estimated TPR and FDR 
based on one hundred  replications.
We also report the predictive root mean squared error defined by ${\rm RMSE}^2={\mathbb E} \{ ({\bf x}_{\rm new}^{\rm T} {\boldsymbol \beta}^* - {\bf x}_{\rm new}^{\rm T} \hat {\boldsymbol \beta})^2\}/{\rm snr}$; 
here the expectation is taken over new predictive locations ${\bf x}_{\rm new}$ and training sets.
Looking at TPR and FDR, the high complexity of CV1se  and the low complexity of SS and $\sqrt{{\rm lasso}}$ are again observed.
Looking at RMSE, QUT$_{\rm lasso}$ often performs best thanks to a good sparse model before fitting by MLE.
Finally, QUT$_{\rm lasso}$ and GIC are comparable in terms of RMSE, but QUT$_{\rm lasso}$ often has a better compromise between TPR and FDR.


%

\subsection{Implementation details}
\label{sct:GPBimplement}



Assuming a Gaussian distribution,  the zero-thresholding function~\eqref{cor:zerothresholdingfctlassoGLM} yields  the null-thresholding statistic 
$$
\Lambda=\| X^{\rm T} (I-P_{X_0}) {\bf Y}_{0}   \|_{\infty},
$$
where $P_{X_0}$ is the orthogonal projection onto the range of $X_0$ and the null model is ${\bf Y}_0 \sim {\rm N}(X_0 \intvec, \sigma^2 I)$. 
Since $\Lambda$  is an ancillary statistic for $\intvec$, the quantile universal threshold  can equivalently be defined as
$\lambda^{{\rm QUT}}=\sigma \lambda_Z$, $\lambda_Z$ being the upper $\alpha$-quantile of
$\Lambda_Z=\|X^{\rm T} (I-P_{X_0}) {\bf Z} \|_\infty$,
where ${\bf Z} \sim {\rm N}({\boldsymbol 0}, I_N)$.
Alike other criteria such as SURE, AIC, BIC and GIC, an estimate of $\sigma$ is required; see Appendix~\ref{app:var} for a possible approach.
In contrast, square root lasso's null-thresholding statistic $\Lambda=\| X^{\rm T} (I-P_{X_0}){\bf Y}_0 \|_\infty/\|(I-P_{X_0}){\bf Y}_0 \|_2$ is pivotal
with respect to both $\intvec$ and~$\sigma$, and LAD-lasso's
is pivotal with respect to~$\sigma$ when $P_0=0$. 
%

%

In Poisson and logistic regression, the null-thresholding statistic depends on $\intvec$
which we estimate with the following procedure. First, calculate the MLE of $\intnostarvec$ based on the observed value ${\bf y}$ with the constraint $\hat {{\boldsymbol \beta}}={\bf 0}$
(it is the solution to \eqref{GLMKKT}). Then, solve~\eqref{eq:GLMlassowithA} with the corresponding quantile universal threshold.
Finally, the estimate is $\hat{\intnostarvec}^{\textsc{\tiny MLE}}$ where $(\hat{\intnostarvec}^{\textsc{\tiny MLE}}, \hat{{\boldsymbol \beta}}^{\textsc{\tiny MLE}})$ denotes the MLE based on ${\bf y}$ with covariates selected by the previous procedure.
In Appendix~\ref{app:sens}, we conduct an empirical investigation of the sensitivity of our approach to the estimation of $\intvec$.

The random design setting is the situation where not only the response vector but also the matrix of covariates is random, like all four data sets in Section~\ref{sct:data}. 
To account for the variability due to random design, we define the quantile universal threshold as the upper $\alpha$-quantile of
 $\Lambda=\lambda_0({\bf Y}_{0},[X_0,X])$, with $[X_0,X]$ consisting of independent identically distributed rows.
 If the distribution of $\Lambda$ is unknown, $\lambda^{\rm QUT}$ is easy to compute with a Monte Carlo simulation which requires bootstrapping the rows of $[X_0,X]$. 
Both fixed and random alternatives are implemented in our \textsf{R} package {\bf qut}.

\subsection{Conclusion}

According to Ockham's razor, if two selected models  yield comparable predictive performances, the sparsest should be preferred.
Lasso with QUT tends to be in accordance with this principle  by selecting low complexity models that achieve good predictive performance.
Moreover, a good compromise between high TPR and low FDR is obtained.
A phase transition in variable screening  corroborates these results in Appendix~\ref{app:LOI}.
In comparison with stability selection, 
QUT is better in two ways: first, it offers a better compromise between low complexity and good predictive performance; second, not being based on resampling,
it is faster and its output is not random. 
Finally, we observe that square root lasso has difficulties detecting significant variables, and its  predictive performance is consequently not as good has that of lasso.

\normalsize

\section{Acknowledgements}

We thank Julie Josse for interesting discussions.
The authors from the University of Geneva are supported by the Swiss National Science Foundation.

\appendix

\section{Proofs}
\subsection{Proof of Lemma~\ref{uniquefit}} \label{app:prooflemma}
It follows from the strict convexity of $b$ on $\Theta$ and the convexity of $f\left( \vect{\beta}_0, \vect{\beta} \right)=\| {\boldsymbol \beta} \|_1$ on $\mathcal F$ that the objective function in~\eqref{eq:GLMlassowithA} is convex on $\mathcal{F}$. The solution set is thus convex.

Assume there exists two solutions $( {\hat {{\boldsymbol \beta}_0}^{(1)}}_{\lambda}, \hat{\boldsymbol \beta}_\lambda^{(1)} )$ and $( \hat {{\boldsymbol \beta}_0}_{\lambda}^{(2)}, \hat{\boldsymbol \beta}_\lambda^{(2)} )$ such that $X_0\hat {{\boldsymbol \beta}_0}_{\lambda}^{(1)} + X \hat{\boldsymbol \beta}_\lambda^{(1)} \neq X_0\hat {{\boldsymbol \beta}_0}_{\lambda}^{(2)} + X \hat{\boldsymbol \beta}_\lambda^{(2)}$. Because the solution set is convex, $( \hat {{\boldsymbol \beta}_0}_{\lambda}^{(3)}, \hat{\boldsymbol \beta}_\lambda^{(3)} ):=\delta ( \hat {{\boldsymbol \beta}_0}_{\lambda}^{(1)}, \hat{\boldsymbol \beta}_\lambda^{(1)} ) + \left( 1-\delta\right)( \hat {{\boldsymbol \beta}_0}_{\lambda}^{(2)}, \hat{\boldsymbol \beta}_\lambda^{(2)} ) $ is a solution for any $0<\delta<1$.
However,
\begin{equation*}
- \ell \left( \hat {{\boldsymbol \beta}_0}_{\lambda}^{(3)}, \hat{\boldsymbol \beta}_\lambda^{(3)} ; {\bf y} \right)  + \lambda \| \hat{\boldsymbol \beta}_\lambda^{(3)} \|_1 < m,
\end{equation*}
where $m$ denotes the minimum value of the objective function and the strict inequality follows from the strict convexity of $b$ and the convexity of  $f\left( {\boldsymbol \beta}_0, {\boldsymbol \beta} \right)=\| {\boldsymbol \beta} \|_1$. In other words, $( \hat {{\boldsymbol \beta}_0}_{\lambda}^{(3)}, \hat{\boldsymbol \beta}_\lambda^{(3)} )$ is not in the solution set, a contradiction.
\subsection{Proof of Theorem~\ref{thm:KKT}} \label{app:proof}
Minimizing \eqref{eq:GLMlassowithA}
over $\mathcal F$ is equivalent to minimizing
\begin{equation*}
f\left( \intnostarvec, {\boldsymbol \beta} \right)=\begin{cases} - \ell \left( \intnostarvec,{\boldsymbol \beta}; {\bf y} \right)  + \lambda \| {\boldsymbol \beta} \|_1 &\mbox{if } \left( \intnostarvec, {\boldsymbol \beta} \right) \in \mathcal{F}, \\ 
+ \infty & \mbox{if } \left( \intnostarvec, {\boldsymbol \beta} \right) \notin \mathcal{F}, \end{cases}
\end{equation*}
over all of $\mathbb{R}^{P_0+P}$. Assuming $f$ is convex, a given point $(  \intnostarvec,  {\boldsymbol \beta})$ belongs to the minimum set of $f$
if and only if $\bf 0$ is a subgradient of $f$ at $(  \intnostarvec,  {\boldsymbol \beta} )$. This is equivalent to
\begin{equation*}
\begin{cases}
X_0   \intnostarvec + X {\boldsymbol \beta}  \in {\Theta}^N,\\
{X_0}^{\rm T} ( {\bf y} - b^{'} \left( X_0   \intnostarvec + X   {\boldsymbol \beta}\right))=0 , \\
X^{\rm T} ( {\bf y} - b^{'} \left( X_0   \intnostarvec + X  {\boldsymbol \beta}  \right)  )  =\lambda {\boldsymbol \gamma},
\end{cases}
\end{equation*}
for some ${\boldsymbol \gamma} \in \mathbb{R}^P$ such that
\begin{equation*}
\gamma_p \in \begin{cases} \left\lbrace \operatorname{sign} ( {\beta}_p)\right\rbrace  &\mbox{if }  {\beta}_p \neq 0, \\ \left[ -1,1 \right]  &\mbox{if }  {\beta}_p = 0, \end{cases}, \quad p=1, \ldots, P.
\end{equation*} 
Setting $(   \intnostarvec,   {\boldsymbol \beta})=(  \hat \intnostarvec_\lambda, {\bf 0} )$ and assuming $b$ is convex, the assertion in Theorem \ref{thm:KKT} follows.

\section{Variance estimation in linear models}
\label{app:var}

When $P>N$ in~\eqref{eq:regression} ($P_0=0$ is assumed for simplicity), constructing a reliable estimator for $\sigma^2$ is a challenging task and several estimators have been proposed. \citet{ReidTibshFriedarchiv14} consider an estimator of the form
\begin{equation}\label{eq:reid}
\hat \sigma^2 = \frac{1}{N -{\hat s}_\lambda} \| {\bf Y} - X  \hat {{\boldsymbol \beta}}_\lambda \|_2^2,
\end{equation}
where $\hat {\boldsymbol \beta}_{\lambda}$ is the lasso estimator tuned with cross-validation and ${\hat s}_\lambda$ denotes the number of estimated nonzero entries.
\citet{Fan:Guo:Hao:2012} propose refitted cross-validation (RCV). The data set is split into two equal parts, $(X^{(1)}, {\bf Y}^{(1)})$ and $(X^{(2)}, {\bf Y}^{(2)})$. On each part, a model selection procedure is applied resulting in two different sets of nonzero indices ${\hat M}_1$, ${\hat M}_2$ with respective cardinality ${\hat m}_1$ and ${\hat m}_2$. This allows to compute
\begin{equation*}
\hat \sigma_1^2=\dfrac{1}{N/2-{\hat m}_1} \| (I - P_{X^{(2)}_{{\hat M}_1}}) {\bf Y}^{(2)}\|_2^2 \quad \mbox{and} \quad \hat \sigma_2^2=\dfrac{1}{N/2-{\hat m}_2}  \| (I - P_{X^{(1)}_{{\hat M}_2}}) {\bf Y}^{(1)} \|_2^2,
\end{equation*}
where $P_{X^{(i)}_{{\hat M}_j}}$ is the orthogonal projection matrix onto the range of the submatrix of $X^{(i)}$ with columns indexed by ${\hat M}_j$.
Finally, the RCV estimator is defined as
\begin{equation}\label{eq:sigma12}
\hat \sigma_{{\rm RCV}}^2:=\dfrac{\hat \sigma_1^2 + \hat \sigma_2^2}{2}.
\end{equation}
Consistency and asymptotic normality hold under some regularity assumptions. In practice, the lasso tuned with cross-validation is applied in the first stage.

\begin{figure}[!ht]
  \begin{center}
  \includegraphics[width=5in]{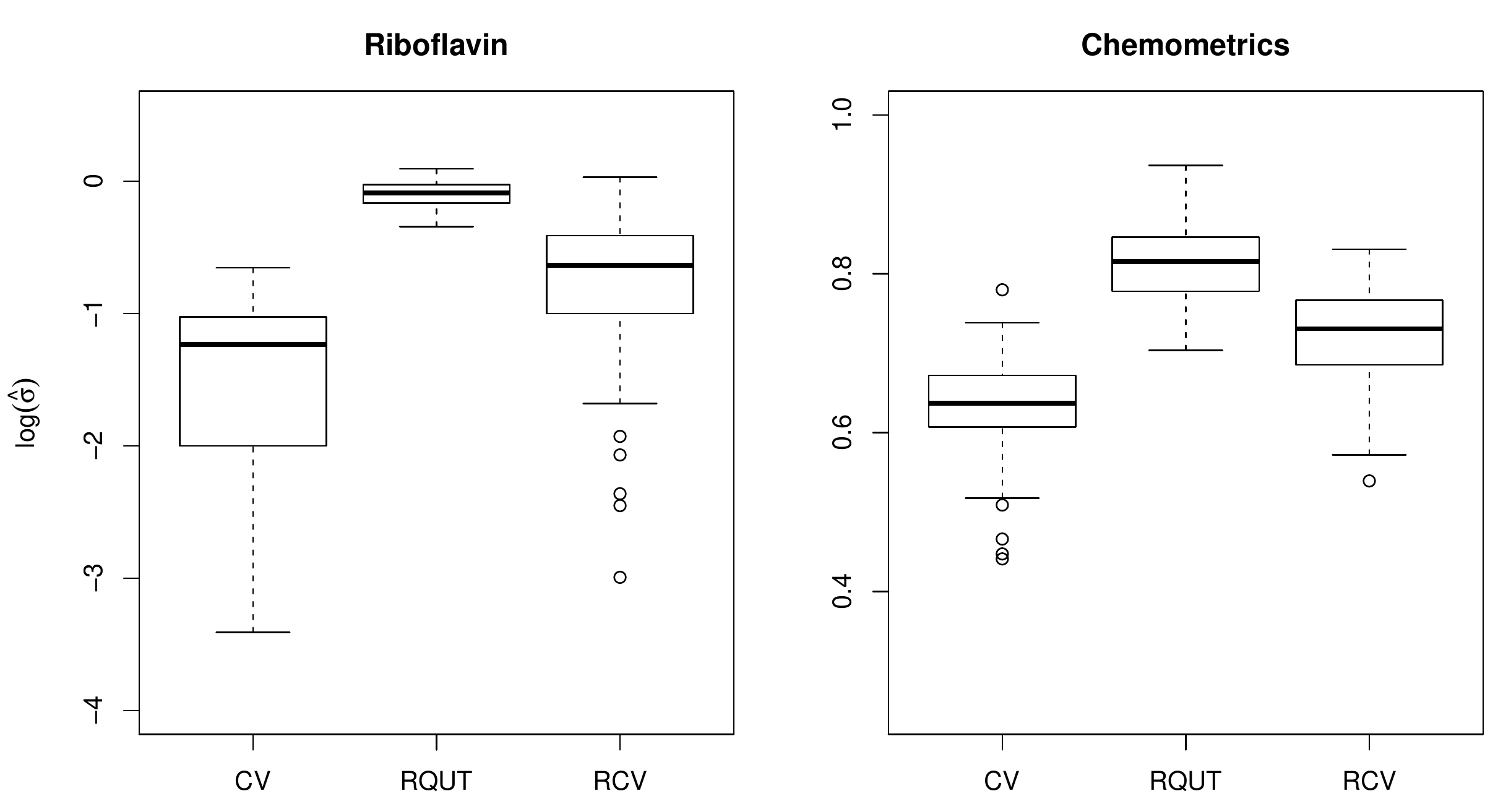}
  \caption{Results of Monte Carlo simulation based on {\tt riboflavin} and {\tt chemometrics} data of Section~\ref{sct:data} for the estimation of $\sigma$ with  cross-validation (CV) defined in \eqref{eq:reid},
  refitted QUT defined in~\eqref{eq:sigma2QUT} and refitted cross-validation (RCV) defined in~\eqref{eq:sigma12}.
   \label{fig:sigmasrealdata}}
  \end{center}
\end{figure}

We propose a new estimator of $\sigma^2$, refitted QUT, which is defined as 
\begin{equation}  \label{eq:sigma2QUT}
\hat \sigma^2_{{\rm QUT}}:=\argmin_{\sigma^2>0} \, \left |\sigma^2 - \hat \sigma_{{\rm RCV}}^2 (\sigma^2) \right |,
\end{equation}
where $\hat \sigma_{{\rm RCV}}^2 (\sigma^2)$ is the RCV estimate with the lasso tuned with $\lambda^{\rm QUT}(\sigma^2)$.
Figure~\ref{fig:sigmasrealdata} shows boxplots of the three estimators of variance applied to the Gaussian data of Section~\ref{sct:data}.
Refitted QUT has smallest variability and seems slightly more conservative than CV and RCV.

\section{Sensitivity study}
\label{app:sens}

\begin{figure}[!ht]
  \begin{center}
  \includegraphics[width=6.2in]{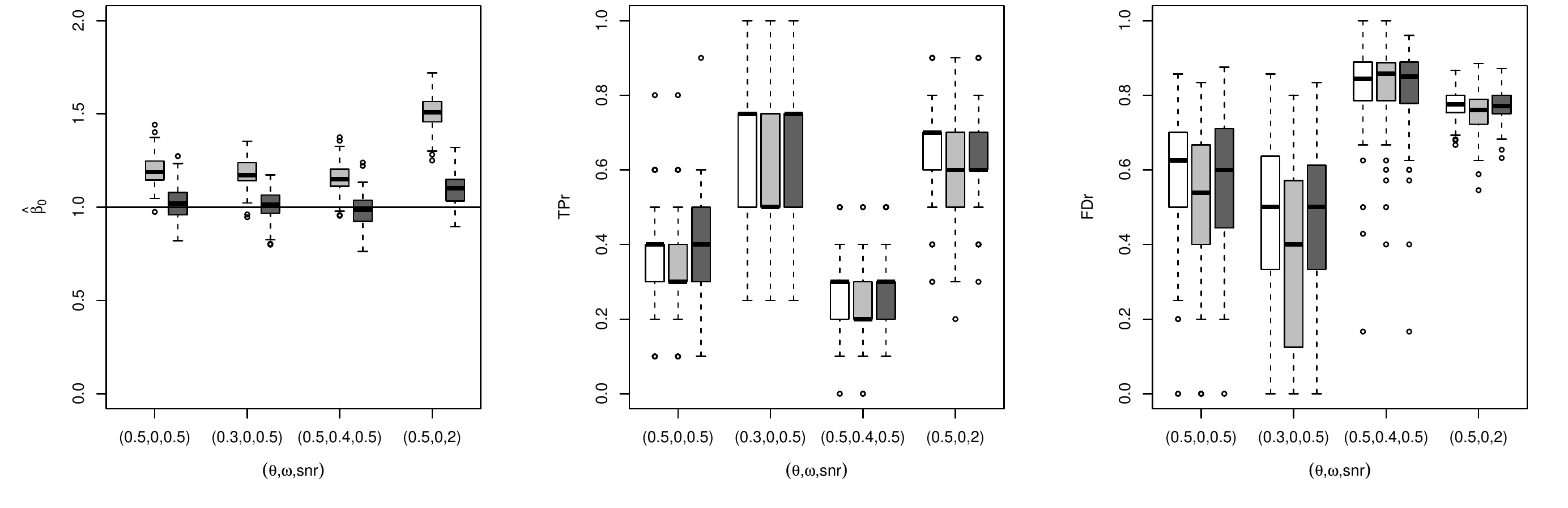}
  \caption{Estimation of $\intscal=1$ (left) and its effect on  TPr (middle) and FDr (right).
  White, light grey and dark grey boxplots correspond respectively to the oracle estimator $\hat \beta_0=1$, initial step and final step of our estimation procedure. 
   \label{fig:alphastar}}
  \end{center}
\end{figure}

As noted in Section~\ref{sct:GPBimplement}, the null-thresholding statistic and therefore the quantile universal threshold are functions of the unknown intercept $\intscal$.
In Figure~\ref{fig:alphastar}, we empirically investigate the sensitivity of our method to the estimation of  $\intscal=1$ on the Poisson distributed data of Section~\ref{subsct:GaussianSimu}.
On the left panel, estimation of $\intscal$ (dark grey) described at the end of Section~\ref{sct:GPBimplement} has low bias.
Moreover we observe the relative median insensitivity of  TPr and FDr to the estimate.

\section{Phase transition property}
\label{app:LOI}
%

We now investigate the variable screening property and observe a phase transition. Given a thresholding estimator,
if several tuning parameter values yield $\hat {\cal S}_\lambda$ containing the true support ${\cal S}^*$, the smallest estimated model can be of interest since it minimizes the FDr.
We call it the optimal inclusive model.
This leads to the definition of the oracle inclusive rate which measures its cardinality relative to the estimated support.


\begin{definition}
Assume ${\cal S}^*\neq \emptyset$ and let $s_{\rm min}:=\min_\lambda \, \{ |\hat {\cal S}_\lambda|: \,  \hat {\cal S}_\lambda \supseteq {\cal S}^*\}$ if it exists. Let ${\hat s}_\lambda := |\hat {\cal S}_\lambda|$ be the cardinality of $\hat {\cal S}_\lambda$.
The oracle inclusive rate (OIR) is defined as ${\mathbb E}[ {\rm OIr}]$, where
\begin{equation*} 
   {\rm OIr}:=\left \{ \begin{array}{ll}
                       \dfrac{s_{\rm min}}{{\hat s}_\lambda} & {\mbox if}\  \hat {\cal S}_\lambda \supseteq {\cal S}^*,\\
                       0 & {\mbox otherwise.}
                      \end{array}
   \right .
 \end{equation*}
\end{definition}

\noindent Models with ${\rm OIr} \neq 0$ have ${\rm TPr}=1$, whereas those with ${\rm OIr}=1$ have minimum ${\rm FDr}$ amongst all models with ${\rm TPr}=1$. Moreover, ${\rm OIR} \leq \mathbb{P}( \hat {\cal S}_\lambda \supseteq {\cal S}^* )$.
A small OIr results from a complex model containing ${\cal S}^*$, whereas a null OIr results from  $\hat {\cal S}_\lambda \supsetneq {\cal S}^*$. The latter could be due to a simplistic model or the variable screening property being unachievable, in which case $s_{\rm min}$ does not exist.

%
%

We extend the simulation of \citet{preciseunder10} in compressed sensing  to  model~\eqref{eq:regression} with unit noise variance assumed to be known. The entries of the $N \times P$ $X$ matrix are assumed to be i.i.d.~standard Gaussian.
We set $P = 1600$ and vary the number of rows $N \in  \{ 160,  320,  480,  640,  800,  960, 1120, 1280, 1440 \}$ as well as the cardinality of the support of ${\boldsymbol \beta}^*$, $s^* \in \{1,\ldots,N\}$. Nonzero entries are set to ten. One hundred predictor matrices $X$ and responses ${\bf y}$ are generated for each pair $(N,s^*)$.

\begin{figure}[!ht]
   \begin{center}
   \includegraphics[height=6cm]{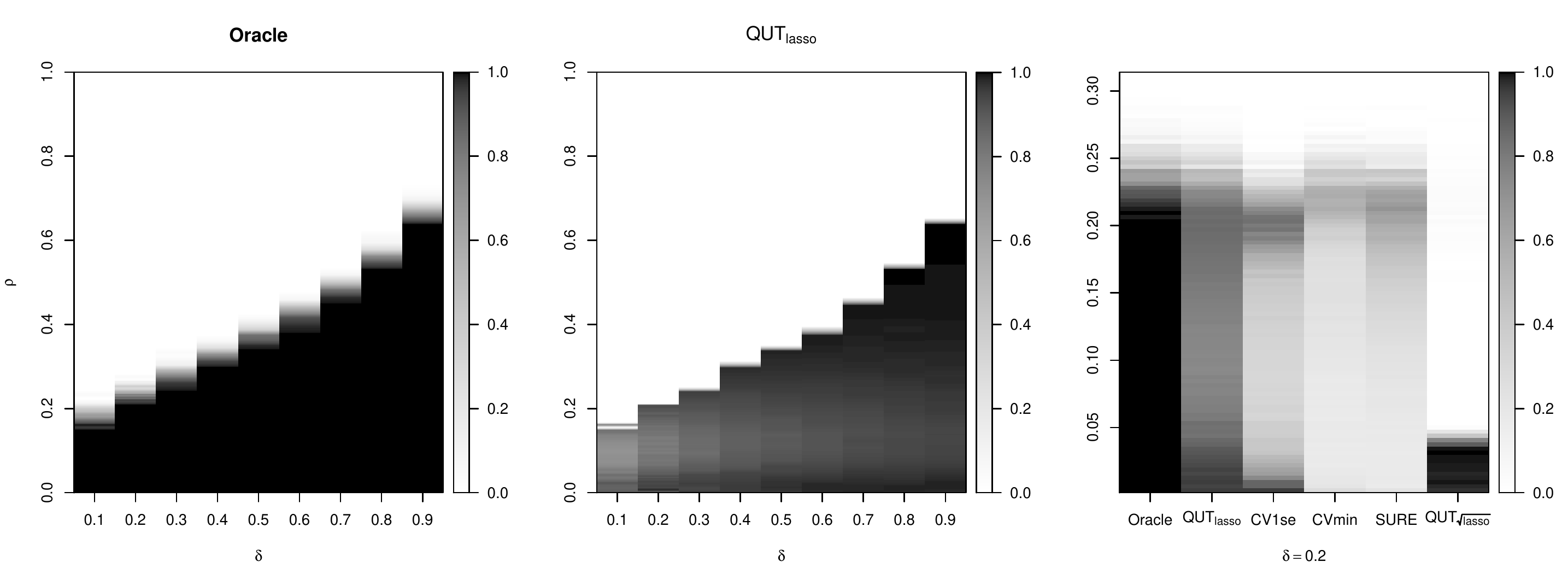}
   \caption{
   Estimated OIR of the oracle lasso selection rule (left) and QUT (middle) as a function of $(\delta,\rho)=(N/P,s^*/N)$. The right panel contains the estimated OIR of several selection rules for a fixed $\delta=0.2$.
   \label{fig:LOIdelta}}
   \end{center}
 \end{figure}

On the left panel of Figure~\ref{fig:LOIdelta}, we report OIR for the oracle lasso selection rule which retains the optimal inclusive model if it exists.
Values are plotted as a function of  $\delta=N/P$, the undersampling factor, and of  $\rho=s^*/N$, the sparsity factor.
On the middle and right panel, we report OIR for QUT$_{\rm lasso}$ along other methodologies as well as QUT$_{\sqrt{\text{lasso}}}$.
The following interesting behaviors are observed:
\begin{itemize}
 \item Phase transition of Oracle and QUT. Two regions can be clearly distinguished: a high OIR region due to a selected model containing few covariates outside the optimal model and a zero OIR region in which $s_{\rm min}$ does not exist. The change between these regions is abrupt, as observed in compressed sensing.
 \item Near oracle performance of QUT. Comparing the left and middle panels, the performance of QUT is nearly as good as that of the oracle selection rule, with the phase transition occurring at similar values of $\rho$.
 \item Low complexity of QUT$_{\rm lasso}$. Comparing several rules on the right panel, QUT has a high OIR. Moreover, CVmin has lower OIR than CV1se and is comparable to SURE. The low OIR of the three latter selection rules is due to the complexity of their selected model. This goes along the fact they are prediction-based methodologies whereas QUT aims at a good identification of the parameters.
 \item Low OIR of QUT$_{\sqrt{\text{lasso}}}$. This could be due to the fact that $\sqrt{\text{lasso}}$ requires stronger conditions than lasso with a known variance to achieve variable screening \citep{BCW11}.
 Considering its zero-thresholding function, not only its numerator increases as the model deviates from the null model (as lasso),
 but also its denominator, making screening harder to reach with $\alpha=0.05$.
 \end{itemize}
 
\bibliographystyle{plainnat}
\bibliography{article_bis}


\newpage 

\end{document}